\documentclass[letterpaper]{article} 
\usepackage{aaai2026}  
\usepackage{times}  
\usepackage{helvet}  
\usepackage{courier}  
\usepackage[hyphens]{url}  
\usepackage{graphicx} 
\usepackage{natbib}  
\usepackage{caption} 
\usepackage{algorithm}
\usepackage{algorithmic}

\usepackage{amsmath, amssymb}
\usepackage{array}
\usepackage{booktabs}
\usepackage{multirow}
\usepackage{xcolor}
\newcommand{\answerYes}[1]{\textcolor{blue}{#1}} 
 
\newcommand{\answerNA}[1]{\textcolor{gray}{#1}} 

\newcommand{\dataset}[1]{\mathcal{D}_{\text{#1}}}

\usepackage{enumitem}

\newcommand{\para}[1]{\noindent\textbf{\textit{#1}~}}

\usepackage{newfloat}
\usepackage{listings}
\DeclareCaptionStyle{ruled}{labelfont=normalfont,labelsep=colon,strut=off} 
\floatstyle{ruled}
\newfloat{listing}{tb}{lst}{}
\floatname{listing}{Listing}

\title{Uncovering the Internet’s Hidden Values: An Empirical Study of Desirable Behavior Using Highly-Upvoted Content on Reddit}

\author{
Agam Goyal, Charlotte Lambert, Yoshee Jain, Eshwar Chandrasekharan
}
\affiliations{
Siebel School of Computing and Data Science\\ University of Illinois Urbana-Champaign, Urbana, IL, USA\\
\{agamg2, cjl8, yosheej2, eshwar\}@illinois.edu}

\begin{document}

\maketitle

\begin{abstract}
A major task for moderators of online spaces is norm-setting, essentially creating shared norms for user behavior in their communities. Platform design principles emphasize the importance of highlighting norm-adhering examples and explicitly stating community norms. However, norms and values vary between communities and go beyond content-level attributes, making it challenging for platforms and researchers to provide automated ways to identify desirable behavior to be highlighted. Current automated approaches to detect desirability are limited to measures of prosocial behavior, but we do not know whether these measures fully capture the spectrum of what communities value. In this paper, we use upvotes, which express community approval, as a proxy for desirability and examine 16,000 highly-upvoted comments across 80 popular sub-communities on Reddit. Using a large language model, we extract values from these comments across two years (2016 and 2022) and compile 64 and 72 \textit{macro}, \textit{meso}, and \textit{micro} values for 2016 and 2022 respectively, based on their frequency across communities. Furthermore, we find that existing computational models for measuring prosociality were inadequate to capture on average 82\% of the values we extracted. Finally, we show that our approach can not only extract most of the qualitatively-identified values from prior taxonomies, but also uncover new values that are actually encouraged in practice. 
Our findings highlight the need for nuanced models of desirability that go beyond preexisting prosocial measures. This work has implications for improving moderator understanding of their community values and provides a framework that can supplement qualitative approaches with larger-scale content analyses.
\end{abstract}

%

\section{Introduction}

Online communities have become a significant medium for human interaction, with users engaging across diverse platforms and content~\cite{williamson2020us}. These interactions give rise to a range of behaviors, some of which result in negative outcomes, while others lead to positive ones, both of which can impact users and platforms~\cite{halfaker_dont_2011, cunha_warm_2017, chandrasekharan_quarantined_2022,cook_awe_2022,verma_examining_2022}. Prior research has concentrated on detecting and examining undesirable behaviors including toxicity~\cite{pavlopoulos2017deeper, chandrasekharan_crossmod_2019}, hate speech~\cite{chandrasekharan_you_2017}, personal attacks~\cite{warner2012detecting}, and employed empirical approaches to uncover norm violations at scale~\cite{chandrasekharan_internets_2018} and determine the prevalence of antisocial behavior~\cite{park_measuring_2022}. Additional approaches include predicting conversational outcomes based on initial comments~\cite{zhang_conversations_2018, bao_conversations_2021}, analyzing the structure of toxic conversations~\cite{ niculae_conversational_2016}, and assessing the \textit{resilience} (i.e., ability to bounce back) of online conversations after adverse events~\cite{lambert_conversational_2022}.

Recently, this focus on addressing undesirable behavior---via abuse detection and punitive actions---has shifted towards exploring complementary strategies to detect and encourage desirable outcomes~\cite{jurgens_just_2019}. These strategies aim to promote the kind of activity that platforms seek to encourage in order to improve individual user well-being and overall community health. \citet{bao_conversations_2021} analyzed the utility of conversational features for the prediction of \textit{prosocial} outcomes in online
interactions by introducing theory-inspired metrics. \citet{weld_making_2024} conducted a survey of Reddit users to develop a comprehensive taxonomy of \textit{community values} that users find desirable to foster better communities. Relatedly, \citet{lambert_positive_2024} surveyed Reddit moderators and constructed taxonomies capturing what moderators want to encourage and what actions they take to positively reinforce desirable contributions.

Despite these efforts, a deeper understanding of the specific attributes in content and behavior that different communities find desirable and encourage in practice remains underdeveloped. Although prior work~\cite{chandrasekharan_crossmod_2019} has used comment removals to study ``social norms,'' implicit guidelines that shape community interactions and determine what behaviors are acceptable~\cite{Bicchieri2011-BICSN}, there is a gap in our understanding of \textit{social values}. Social values are the fundamental principles and ideals of a community serving as the foundation from which norms emerge and gain power through enforcement~\cite{mcclintock1978social}.

Similar to social norms, identifying social values at scale is challenging due to their emergent nature and the considerable variation across communities. These variations are influenced by factors such as governing norms, the sensitivity of interaction topics, and target audiences~\cite{jurgens_just_2019}. 
Analyzing these values requires large-scale qualitative assessments, as traditional computational approaches like topic modeling~\cite{blei2003latent} tend to focus on the content of conversations rather than the extraction of underlying values.
Additionally, the extent to which existing prosociality metrics~\cite{bao_conversations_2021} can explain the types of actions and content valued by communities is unclear. Moreover, prior survey studies have highlighted that prosociality is indeed desired by community moderators~\cite{lambert_positive_2024}, and our goal is to extend this line of work by examining how prevalent these attributes are across multiple communities along with uncovering the additional attributes of desirable content.

In this paper, we present a framework to computationally extract the community values being encouraged in practice by Reddit communities, called subreddits. 
We focus on upvotes
as a proxy for the desirability of content. 
Key reasons for using upvotes are that (1) upvotes are the default way to signal desirability, are signals that indicate the same treatment across communities, and are freely available to use for all users on the platform unlike other signals like `awards'; and (2) recent work~\cite{lambert_positive_2024,10.1145/3706598.3713830} has shown that upvotes are being used by moderators to enforce certain values in practice, and receiving upvotes encourages users to post more desirable content in the future.

We base our extraction of community values on 16,000 highly-upvoted comments from 80 different Reddit communities.
First, we develop a simple approach powered by large language models (LLMs) to extract and categorize values as macro, meso, or micro based on their prevalence across communities. 
In order to validate the robustness of our approach, we perform our analysis on data from two years, 2016 and 2022.
Next, we evaluate whether existing measures of desirable behavior (e.g., prosociality~\cite{bao_conversations_2021}) from prior work are sufficient to explain the likelihood of a comment being highly upvoted by its community, and can adequately capture the community values we extract in our work. We then analyze how our computationally-extracted values compare to existing taxonomies of desirable behavior~\cite{lambert_positive_2024} and community values~\cite{weld_making_2024}.

Specifically, we address the following research questions:
\begin{enumerate}[leftmargin=1.0cm]
    \setlength{\itemsep}{0pt}
    \item[\textbf{RQ1:}] What types of behavior and content are highly upvoted (i.e., encouraged) within subreddits? How do these values overlap and vary across communities?
    \item[\textbf{RQ2:}] How well do existing computational approaches to quantify desirability explain what gets highly upvoted?
    \item[\textbf{RQ3:}] How do existing survey-based taxonomies of desirable behavior intersect with the values we extract?
\end{enumerate}

\para{Summary of Findings:} 
Using our large language model (LLM)-based approach, we analyzed popular content across 80 popular Reddit communities for two years, 2016 and 2022. Our analysis revealed 64 distinct values in 2016 and 72 values in 2022, spanning macro, meso, and micro scales. We find that communities reward prosocial behavior to varying degrees and that existing prosociality measures alone are insufficient to quantify desirability, with current models capturing only 14 out of the 64 values in 2016 and 10 out of the 72 values in 2022 with at least 25\% recall. Finally, we show that our extracted values align closely with existing survey-based taxonomies of desirable behavior and present deeper insights into values at meso and micro levels. This work has implications for researchers studying desirable behavior online and for platforms aiming to develop automated moderation tools that can detect and encourage desirable behavior. We highlight the need to broaden definitions of prosociality and desirability within online communities, and lay the foundation for future work on computational models that align with community-specific values.

\section{Related Work}

Understanding, quantifying, and predicting user behaviors and conversational outcomes within online communities has been a growing area of research. \citet{chandrasekharan_internets_2018} conducted a large-scale empirical study on norm violations across 100 subreddits to identify emergent norms, while \citet{park_measuring_2022} examined the prevalence of antisocial behavior within these communities. Various studies have also looked at toxicity and hate speech~\cite{chandrasekharan_you_2017, kumar2017antisocial, chandrasekharan_crossmod_2019}, mental well-being~\cite{saha_prevalence_2019, saha_causal_2020, saha_social_2022,shimgekar2025detecting,pal2026hidden}, moderation outcomes~\cite{chandrasekharan_quarantined_2022, jhaver_evaluating_2021}, and forecasting conversational outcomes~\cite{liu2018forecasting, chang_trajectories_2019, bao_conversations_2021, lambert_conversational_2022}. \textit{Our work aims to enhance the understanding of the kinds of behavior and content that online communities find desirable.}

\para{Prior Work on Desirable Behavior Online:} \citet{jurgens_just_2019} highlighted the need to expand the focus of computational approaches to handle online abuse to be more proactive (i.e., not wait until the bad behavior has occurred). Moreover, the absence of abuse does not necessarily imply that a community is thriving and healthy~\cite{bao_conversations_2021}.
As a result, we need to complement current moderation strategies with more proactive approaches to incentivize and encourage normative-behavior (e.g., via norm-setting~\cite{grimmelmann_virtues_2015}, rewards or positive feedback~\cite{kraut_building_2011, kiesler_regulating_2012}, highlighting or increasing the visibility of desirable behavior~\cite{diakopoulos_editors_2015, choi_creator_2024}).
Researchers have developed computational approaches to detect different types of positive behavior like politeness~\cite{danescu-niculescu-mizil_computational_2013}, support~\cite{wang2018s}, empathy~\cite{zhou2020condolence}, constructiveness~\cite{kolhatkar2017constructive}, and prosociality~\cite{bao_conversations_2021}. 
\citet{bao_conversations_2021} applied social psychology theories to operationalize supportiveness~\cite{wang_its_2018}, agreement~\cite{brown1991self}, and politeness~\cite{danescu-niculescu-mizil_computational_2013} as three key measures of prosociality. These measures 
were used to predict prosocial outcomes in conversations. \citet{lambert_conversational_2022} similarly applied these measures to 
extract 
conversation features that
result in prosocial outcomes following norm violations.

In addition to computational approaches, researchers have also employed human-centered approaches to explore what community-members and moderators value. \citet{weld_making_2024} surveyed users across Reddit to construct a taxonomy of community values. \citet{lambert_positive_2024} similarly compiled a taxonomy of attributes Reddit moderators want to encourage in their communities and found that many of them took explicit action to positively reinforce those attributes. Importantly, the most common attribute moderators found desirable was being prosocial. These analyses present valuable insights into user and moderator perspectives on what is desirable, but there is 
still work to be done to identify what values are being encouraged in practice.
\textit{This paper aims to examine the recall of current computational methods to quantify desirable behavior and augment prior taxonomies of desirable behavior by identifying the values being rewarded by communities in practice.}

\para{LLMs for Qualitative Analysis:} Large language models (LLMs) are increasingly being applied for qualitative analysis in computational social science research, and have shown both potential and challenges~\cite{ziems2024can}. \citet{gilardi2023chatgpt} demonstrated that models like ChatGPT can outperform human crowdworkers on text annotation tasks, suggesting that LLMs hold promise for automating labor-intensive processes, and~\citet{bonikowski2022ends} surveyed recent works integrating computational text analysis approaches for social science research. \citet{abdurahman2024perils} explored the opportunities and challenges of using LLMs in psychological research for task automation to expand our understanding of human psychology. 
Notably,~\citet{peters2024large} investigate whether LLMs can infer the psychological dispositions of social media users and show that they can perform at a level similar to that of supervised machine learning models, and~\citet{chew2023llm} also show that ChatGPT can perform deductive coding for content-analysis tasks across multiple domains with agreement levels similar to human annotators. Similarly, \citet{dunivin2024scalable} shows that chain-of-thought reasoning with GPT-4 matches human performance in some qualitative coding tasks in a scalable manner. Related to our work, ~\citet{park2024valuescope} propose a framework leveraging LLMs to quantify
social norms, community preferences, and values within online communities, and \citet{lam2024concept} propose an LLM-based framework to analyze unstructured text using concept induction. \textit{Given this potential for language models to effectively perform annotation tasks, we use them as a tool to extract community values.}

\section{Data Collection Pipeline}



Our analyses rely on data collected from 80 popular Reddit communities (or subreddits).
Reddit is an apt platform to study variations in notions of desirable behavior due to the wide range of topics covered by different subreddits and the overall size of Reddit's user base.

For this study, we began with a dataset of 100 subreddits from which \citet{chandrasekharan_crossmod_2019} originally compiled comment removal data between May 10, 2016 and February 4, 2017. This dataset was further extended by \citet{lambert_conversational_2022} to include all comments posted on these 100 subreddits during this period. We refer this dataset as $\dataset{2016}$. To ensure robustness of our overall framework, we also collected data for the same 100 subreddits from the most recent full year for which we had comment data at the time of our study: Jan 1, 2022 to Dec 31, 2022, which we refer to as $\dataset{2022}$.

We then filtered out all comments that were removed by moderators, deleted by the authors themselves, or posted by authors whose accounts were deleted. This filtering avoids comments without text to analyze and maintains privacy for authors who did not want their content visible on the platform. Furthermore, we filtered out comments from accounts that were either known moderation bots or moderators.

Since we are interested in extracting values that would explain why  a comment would be highly upvoted, we label each comment with a binary label of ``low'' and ``high'' \texttt{score} (\# upvotes -- \# downvotes). For each subreddit, we label comments below the $70^{\text{th}}$ percentile of \texttt{score} as ``low'' and above the $95^{\text{th}}$ percentile of \texttt{score} as ``high.'' See Appendix \ref{app:thresholds} for details on the choice of the thresholds.

Next, since desirability of a comment and the values it portrays may be context dependent, we construct \{context, highly-upvoted comment\} pairs where the context is the preceding comment as a reply to which the comment was posted. To ensure enough representation from each subreddit while also keeping the analysis tractable, we aimed to obtain $100$ such pairs for each subreddit across $\dataset{2016}$ and $\dataset{2022}$. Furthermore, in order to ensure that $\dataset{2016}$ and $\dataset{2022}$ are comparable, we randomly sampling these pairs across the data for the particular year to avoid any chronological biases and stop once we find $100$ pairs for each subreddit. However, we found that some of the original 100 subreddits had been banned, and other did not have at least 100 pairs where the comment reached the `high' upvote category at the 95th percentile. This filtering process resulted in our final set of 80 subreddits. Our final study dataset $\dataset{study}$ therefore comprises of 16,000 \{context, highly-upvoted comment\} pairs from $80$ subreddits.

These 80 subreddits vary in size, with subscriber counts ranging from approximately 96K for \texttt{r/AskTrumpSupporters} to 66.6M for \texttt{r/funny} (as of in March 2025). The set of communities is also topically diverse consisting of entertainment, science, question-answering, sports, debate, technology, and humor communities (among others). A complete list of the 80 subreddits can be found in Appendix \ref{app:subreddit-list}.

\section{RQ1: Computationally Extracting Values Across Communities at Scale}

We now introduce our framework for computationally extracting values beyond prosociality that may influence upvoting patterns. Details about our choice of language model and inference parameters are provided in Appendix \ref{app:model-parameters}.

\subsection{Methodology}

\para{Prompting Pipeline:} Since traditional approaches like topic modeling~\cite{blei2003latent} focus on topic-level attributes and are not well-suited to extracting higher-level traits and values from text, we make use of a state-of-the-art large language model (LLM). Specifically, we query GPT-4o~\cite{achiam2023gpt}, providing it with the subreddit description, and \{context, comment\} pairs from $\dataset{study}$, and task it with extracting between one and three keywords that make the comment highly upvoted given the context. We also use chain-of-thought reasoning~\cite{wei2022chain} to elicit explanations from the model for the extracted keywords. See Appendix \ref{sec:prompt-design} for the prompt design and our efforts to ensure reproducibility, and Appendix \ref{app:gemma-replication} for a replication with 4-bit quantized Gemma-3-27B-IT~\cite{team2025gemma}.

The model returns a list of either one, two, or three keywords that capture the traits or values in the comment that may make it upvote-worthy, or ``N/A'' if the model does not consider the comment to be upvote-worthy.
Allowing the model to return ``N/A''s is important to ensure that it is not forced to return a list of traits if it decides the comment should not have been upvoted. 



\para{Overall Framework:} We prompt the LLM to extract up to three keywords for each pair in $\dataset{study}$, separately for the two study years. This process resulted in 249 and 393 unique keywords across all subreddits in $\dataset{2016}$ and $\dataset{2022}$ respectively. 
We observed that some of these keywords were semantically similar (e.g., synonyms, same word-family, etc.). In order to cluster/group keywords that represent similar values and obtain a unique set of values, 
we use a Sentence Transformer model~\cite{reimers-2019-sentence-bert} to embed each keyword to create an embedding pool. Next, on this large set of embeddings, we perform agglomerative clustering to obtain $100$ clusters for each year. 
This initial value of $100$ was determined to ensure that there were neither not too many individual keywords nor clusters which were too dense and had unrelated keywords grouped together. We then generate one-word labels using GPT-4o for the keywords in each cluster. Finally, we manually regroup the values if needed, to modify the clusters to ensure the validity of the final groupings. This grouping leads to a final set of 64 and 72 values for $\dataset{2016}$ and $\dataset{2022}$ respectively, both sets including an ``N/A''. Finally, we perform a value-prevalence analysis to understand how many subreddits show each of the obtained values within the extracted value set.

The result of this framework is a set of computationally-extracted values that are being encouraged (via upvotes) by Reddit communities in practice. While the extracted set of values is not exhaustive for the entire platform, it is a representative sample of highly-occurring traits that community members value across Reddit.


\begin{figure*}
    \centering
    \includegraphics[width=\linewidth]{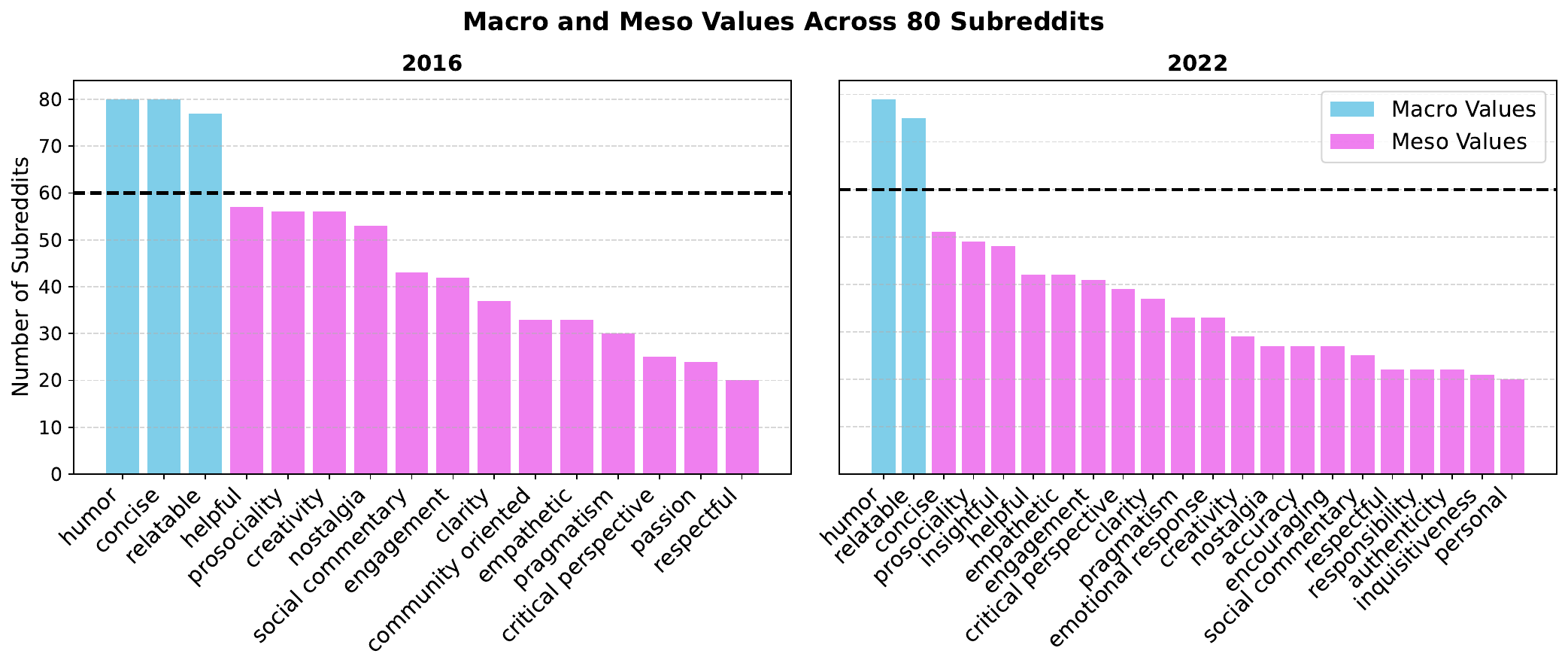}
    \caption{Plot representing the macro and meso values extracted from 80 subreddits in $\dataset{2016}$ (left) and $\dataset{2022}$ (right). Macro values manifest in at least 60 subreddits while meso manifest in between 20 and 60 subreddits. In 2016 we find 3 macro values and 13 meso values, while in 2022 we find 2 macro values and 20 meso values. Dashed line separates macro and meso scales.}
    \label{fig:value_visualization}
\end{figure*}

\para{Reliability of LLM-extracted Values:} Language models have been shown to be well-suited to various zero- and few-shot tasks~\cite{brown2020language, kojima2022large} even though they are not trained on the specific tasks under consideration. They also elicit strong reasoning capabilities when prompted with a chain-of-thought approach~\cite{wei2022chain}. Moreover, prior work has demonstrated that LLMs have familiarity with content from popular Reddit forums~\cite{park2024valuescope, dignan_reddit_2024}.

We further ensure robustness by evaluating the reliability of the LLM-extracted values through a manual annotation task. We randomly sample 160 comments from $\dataset{2022}$ (i.e., two from each subreddit) and have two annotators independently assess these samples to determine whether each of the values extracted by the model are actually exhibited in the comments. The annotators then ensured that any disagreements were due to subjective differences. 

The average accuracy of the labels generated by the LLM, measured as the proportion of the extracted values actually exhibited in the comments was $77\%$. Using Krippendorff's $\alpha$ to measure Inter-Rater Reliability (IRR) between the annotators, we obtained $\alpha=0.61$ which denotes substantial agreement~\cite{hayes2007answering}. The annotators disagreed the most on the value ``relatable'' due to its subjective nature. The IRR measured on all values excluding ``relatable'' is $\alpha=0.73$. These measures highlight the validity of our framework and reliability of the extracted values.

\subsection{Findings}

Similar to the approach \citet{chandrasekharan_internets_2018} took to categorize community norms, we group the extracted community values into macro, meso, and micro values based on their prevalence across subreddits. Values shared by at least 75\% of the 80 subreddits (at least 60) are categorized as \textit{macro values}, those shared by between 25\% and 75\% subreddits are categorized as \textit{meso values}, while those shared by less than 25\% (at most 20) subreddits are categorized as \textit{micro values}. A detailed list of all values is in Appendix \ref{app:value-list}.

\para{Macro values:} Figure \ref{fig:value_visualization} shows macro values extracted from comments in $\dataset{study}$. We find that three macro values emerged from $\dataset{2016}$: \textit{humor}, \textit{relatable}, and \textit{concise}. Two macro values surfaced from $\dataset{2022}$: \textit{humor} and \textit{relatable}. 

A total of 6,785 comments (4,064 from $\dataset{2016}$ and 2,721 from $\dataset{2022}$) were labeled with \textit{humor}, including the following example from \texttt{r/Android} ($\dataset{2022}$):
\begin{quote}
    \textbf{Context:}
    \textit{``Wow, it looks like the Pixel 6... how can this be?!? /s''} 
    
    \textbf{Comment:}
    \textit{``And your gonna love it- Tim Google. Jokes aside i love the Pixel 6 design.''}
\end{quote}

Similarly, a total of 7,752 comments (3,834 from $\dataset{2016}$ and 3,918 from$\dataset{2022}$) were labeled with \textit{relatable}, including the following example from \texttt{r/television} ($\dataset{2022}$):
\begin{quote}
    \textbf{Context:}
    \textit{``When my ex and I broke up and I found out he WAS actually cheating like I suspected. Trust your gut ladies. [...]''}
    
    \textbf{Comment:}
    \textit{``I think gut feelings only go so far. I am immensely paranoid that my partner is cheating/is going to cheat on me, but I know it's down to how I was treated by all of my ex partners.''}
\end{quote}

Finally, a total of 2,276 comments in $\dataset{2016}$ were labeled with \textit{concise}, including the following example from \texttt{r/2007scape} ($\dataset{2016}$):
\begin{quote}
    \textbf{Context:}
    \textit{``Everyone knows there is a viewbot on her stream. [...] Don't give her more attention.''}
    
    \textbf{Comment:}
    \textit{``if only more people understood this simple approach to ignoring people out of existence.''}
\end{quote}

All subreddits in $\dataset{2016}$ valued humorous and concise content while 77 subreddits valued content that was relatable. In $\dataset{2022}$, all subreddits except \texttt{r/legaladvice} valued humorous content while 75 subreddits valued relatable content.

\para{Meso values:} Meso values are less common than macro values, but still appear in between 20 and 60 communities. From Figure \ref{fig:value_visualization} we see that there are 13 and 20 meso values in $\dataset{2016}$ and $\dataset{2022}$ respectively. The meso values that showed up in both samples include \textit{clarity}, \textit{creativity}, \textit{critical} \textit{perspective}, \textit{empathetic}, \textit{engagement}, \textit{helpful}, \textit{nostalgia}, \textit{pragmatism}, \textit{prosociality}, \textit{respectful}, and \textit{social commentary}.

Comments expressing \textit{prosocial} behavior, including politeness, compassion, and positivity, were highly upvoted in 56 and 49 communities in $\dataset{2016}$ and $\dataset{2022}$ respectively. This includes examples such as the following from \texttt{r/changemyview} ($\dataset{2022}$):

\begin{quote}
    \textbf{Context:}
    \textit{``So I am into astrology while dating a person who is not. I do agree that there is some level of people [...] I really enjoy those kinds of people in every area of my life''}
    
    \textbf{Comment:}
    \textit{``Thank you for this perspective. Helps me think about those that hold these beliefs differently $\Delta$''}
\end{quote}

Another common meso value extracted from 53 and 27 communities in $\dataset{2016}$ and $\dataset{2022}$ respectively was \textit{nostalgia}. These comments speak on a personal level and evoke nostalgic responses from other users. Example comments include the following from \texttt{r/2007scape} ($\dataset{2016}$):

\begin{quote}
    \textbf{Context:}
    \textit{``This video makes me so sad, I hate adult life. [RuneScape] was so god damn immersive.''}
    
    \textbf{Comment:}
    \textit{``Yeah I feel you bro. I miss that exciting feeling when playing [RuneScape] as a kid.''}
\end{quote}

\textit{Social commentary} is another meso value which was extracted from 43 and 25 communities in $\dataset{2016}$ and $\dataset{2022}$ respectively. These comments discuss social issues like corruption, justice, fairness, and legal order, among others. Example comments include the following from \texttt{r/SandersForPresident} ($\dataset{2022}$):

\begin{quote}
    \textbf{Context:}
    \textit{``Congress: Urgent stimulus package. Company: Sudden windfall. TV Economists: Inflation? Wall Street: Record Profits!''}
    
    \textbf{Comment:}
    \textit{``That Chips bill bailout is wild. Those companies are making money hand over fist.  Use tariffs, quotas... anything but our tax dollars. And if you must use tax dollars, the government should get a massive block of stock in return.''}
\end{quote}


    

Other meso values specific to $\dataset{2016}$ are \textit{community oriented} and \textit{passion}, and those specific to $\dataset{2022}$ include \textit{accuracy}, \textit{authenticity}, \textit{concise}, \textit{emotional response}, \textit{encouraging}, \textit{inquisitiveness}, \textit{insightful}, \textit{personal}, and \textit{responsibility}.

\para{Micro values:} Micro values appear in only a small set of communities. These communities often share a common trait, such as topic of discussion. Using our framework we extracted a total of $47$ and $49$ micro values from $\dataset{2016}$ and $\dataset{2022}$ respectively. The full set of micro values are available in Appendix \ref{app:value-list}, but we highlight two examples below.

The value \textit{resilience} appeared in seven subreddits in $\dataset{2022}$: \texttt{r/CFB}, \texttt{r/OldSchoolCool}, \texttt{r/churning}, \texttt{r/gifs}, \texttt{r/movies}, \texttt{r/socialism}, and \texttt{r/UpliftingNews}. An example comment exhibiting this value includes the following from \texttt{r/UpliftingNews} ($\dataset{2022}$):

\begin{quote}
    \textbf{Context:}
    \textit{``All I hear is woman lacking proper maternity leave forced to worked until labor''}
    
    \textbf{Comment:}
    \textit{``She was involved in a multi car crash and decided to help someone who was trapped.''}
\end{quote}

\textit{Inclusivity} was another value extracted from fourteen subreddits:  \texttt{r/AskTrumpSupporters}, \texttt{r/wow}, \texttt{r/sex}, \texttt{r/GetMotivated}, \texttt{r/OldSchoolCool}, \texttt{r/books}, \texttt{r/canada}, \texttt{r/changemyview}, \texttt{r/europe}, \texttt{r/dataisbeautiful}, \texttt{r/personalfinance}, \texttt{r/photoshopbattles}, and \texttt{r/nba}. An example comment from \texttt{r/CanadaPolitics} ($\dataset{2022}$):

\begin{quote}
    \textbf{Context:}
    \textit{``who cares. we are all human, most of us get along well and enjoy each others' differences and thats all that matters.''}
    
    \textbf{Comment:}
    \textit{``Agreed. People have migrated since the beginning of time and everyone needs a place to call home. Get to know your neighbours man, there are really great people around you.''}
\end{quote}

Overall, we notice that most individual communities reward content with similar kinds of values in both 2016 and 2022. However, the scale at which values manifest varied, primarily at the meso and micro scales. Moreover, the overall increase in the number of meso and micro values from $\dataset{2016}$ to $\dataset{2022}$ indicates a rise in the range of values that users reward in content over the years, either in individual communities or in a related set of communities. 

\para{Case Study---\textit{r/2007scape} and \textit{r/TwoXChromosomes}:} We now provide descriptive temporal insights using a case study of \textit{r/2007scape}, a subreddit for discussion around the online game Old School RuneScape, and \textit{r/TwoXChromosomes}, a subreddit for both casual and serious discussions centered around women's perspectives. We chose these two communities for their vastly different topics of discussion and community sizes ($\approx$1.2M and $\approx$14M members for \textit{r/2007scape} and \textit{r/TwoXChromosomes} respectively).

We extracted nine values from \textit{r/2007scape} in 2016: ``community oriented,'' ``agreement,'' ``humor,'' ``relatable,'' ``insightful,'' ``prosociality,'' ``concise,'' ``helpful,'' ``nostalgia''. We extracted eleven values from the same community in 2022, where ``expertise,'' ``respectful,'' ``passion,'' and ``critical perspective'' replaced ``agreement'' and ``insightful'' from the set of extracted values from 2016. 
We infer that the values appearing in the highly-upvoted content from 2016 and 2022 represent those more core to the community. Namely, \textit{r/2007scape} seems to value
humorous and relatable content that is concise, prosocial, helpful, community oriented, and nostalgic. 
On the other hand, we observe that the community valued critical perspectives over agreement in 2022 and valued respectful yet passionate content more than it did in 2016. Interestingly, ``N/A''s were not extracted from this community in either year.

For \textit{r/TwoXChromosomes} we see a different temporal trend in which values there was minimal change between values extracted from 2016 and 2022. Those extracted from 2016 include ``personal,'' ``concise,'' ``pragmatism,'' ``humor,'' ``encouraging,'' ``safety,'' ``social commentary,'' ``empathetic,'' ``community oriented,'' ``creativity,'' ``responsibility,'' and ``relatable'', and only change in 2022 is the additional extracted value ``authenticity''. 
This indicates that the extracted values from \textit{r/TwoXChromosomes} may primarily be core community values held across the two time periods.

These observed temporal trends highlight that not only do different communities find distinct values to be desirable, but also that the kind of content valued by the same community can shift over time. The shift in these values across time could depend on factors such as changing community dynamics resulting from rule changes, shift in topics of discussion, or world events. Future work can more deeply explore the reasons for these shifts, and explore similar case studies of temporal shifts using other subreddits.

It is important to note that our current analysis is limited to the extraction of values and descriptive analysis of temporal variations. We are unable to quantify temporal variations (between 2016 and 2022) based on the prevalence of individual values since our data may not capture all possible occurrences of these values due to being a sample of all the comments posted in the entire year. Future work can perform a prevalence analysis to examine temporal variations in values, similar to prior work on prevalence estimation of norm violations on Reddit~\cite{park_measuring_2022}.

\para{When are ``N/A''s returned by the model?} We inspected the 135 instances in  $\dataset{study}$ where the LLM returned ``N/A'' across both years indicating that a comment should not have been highly-upvoted to understand the mismatch between the community and LLM's notions of desirability. The most common themes in these comments
are that they are political, toxic, reference religion, and manifest in the form of controversial opinions, dark humor, or racism.
Prior work ~\cite{rottger2023xstest} has shown that 
reinforcement learning with human feedback (RLHF)~\cite{ouyang2022training} makes the model avoidant of promoting or reflecting positively on content that may be controversial in these ways to maintain safety standards. 

\begin{table}
\small
\sffamily
\centering
\begin{tabular}{l r}
 \textbf{Categories} & \textbf{Frequency} \\
 \midrule
 Offensive/Derogatory Language & 58 \\
 Personal Attack/Harassment & 25 \\
 Off-topic/Irrelevant & 18 \\
 Low Effort/No Value & 15 \\
 Sensitive Topic Handled Insensitively & 10 \\
 Controversial/Divisive Statement & 9 \\
 \bottomrule
\end{tabular}
\caption{Open-coded categories of LLM-generated explanations for comments labeled with ``N/A''.}
\label{tab:open-codes}
\end{table}

We perform manual validation of LLM-generated explanations for ``N/A''s and open-code them into the six categories shown in Table \ref{tab:open-codes}. We find that the model is able to identify instances where the comment does not contribute positively to the discussion, or has derogatory remarks. 
It further identifies content that may be off-topic, or uses personal attacks on the poster of the context. A complete list of subreddits that returned ``N/A''s across the two years in $\dataset{study}$ can be found in Appendix \ref{app:na-subreddits}. 

While the LLM does not consider potentially offensive comments and repetitive comments that do not add meaningfully to the discussion as desirable and worthy of an upvote, the observation that such undesirable behavior was sometimes encouraged (i.e., highly upvoted) within communities further emphasizes the complexities of desirability notions across different online spaces.

\section{RQ2: Examining the Sufficiency of Existing Prosociality Measures in Explaining Upvotes}

Next, we investigate whether existing prosociality measures~\cite{bao_conversations_2021}, namely supportiveness, agreement, and politeness: (a) are sufficient to explain what content receives upvotes on Reddit, and (b) have \textit{recall}---i.e., the ability to identify all data points in a relevant class---for our computationally-extracted values. 
Our focus is to investigate the \textit{recall} of current detection strategies, by examining how well current approaches are able to identify different forms of desirable behavior---using our extracted values.

\begin{figure}[t]
    \centering
    \includegraphics[width=\linewidth]{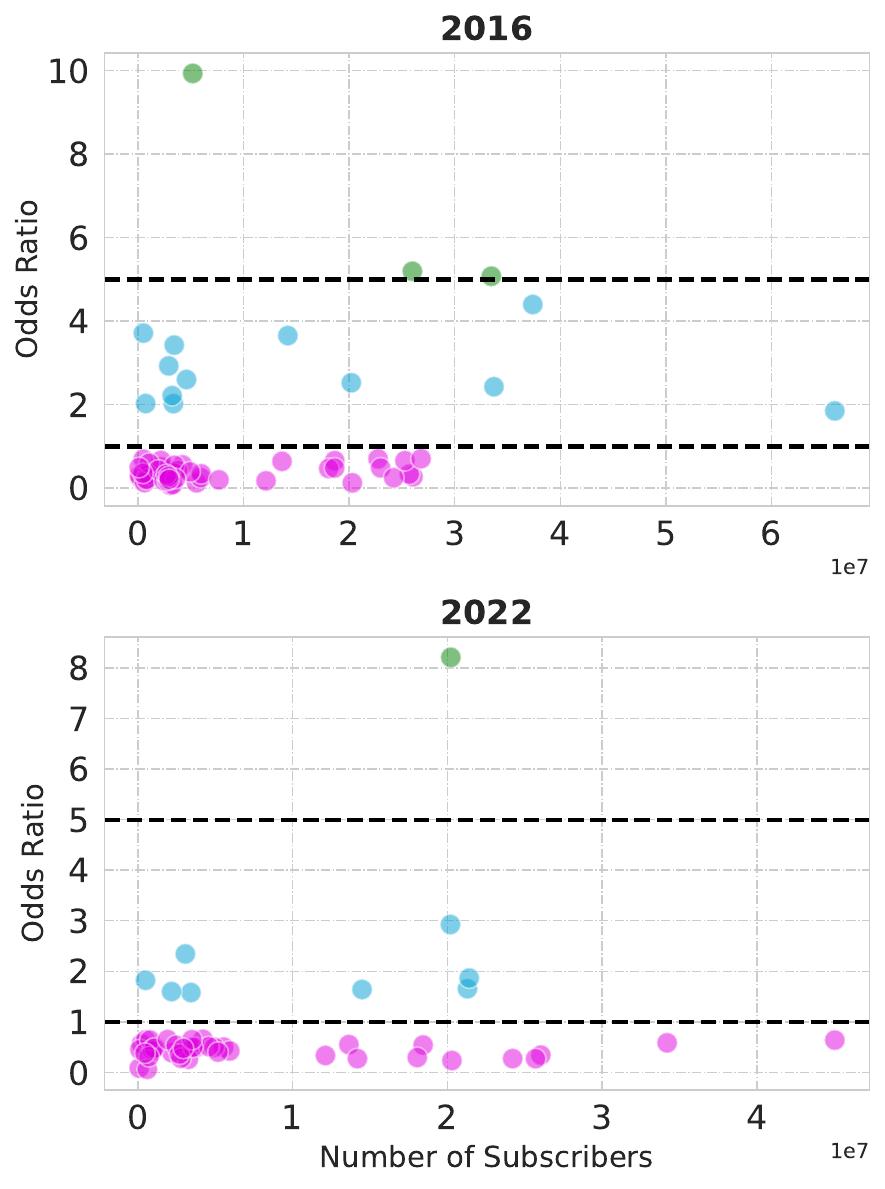}
    \caption{Plot depicting odds ratios by logistic regression analysis on subreddits plotted against the number of subscribers of each subreddit for $\dataset{2016}$ and $\dataset{2022}$. An odds ratio greater than 1 indicates a positive relationship of prosociality measures on likelihood of getting ``high'' upvotes. We see that for the majority of the subreddits in both years (76.3\% in $\dataset{2016}$ and 82\% in $\dataset{2022}$), prosociality alone does not increase the likelihood being highly upvoted. Only subreddits with statistically significant results are plotted.}
    \label{fig:logreg_prosociality}
\end{figure}

\subsection{Sufficiency in Explaining Upvotes}

\para{Methodology:} To conduct our analysis, we randomly sampled 2,500 ``high'' and ``low'' \texttt{score} category comments from the initial stages of our data curation for each subreddit to create two subreddit-specific datasets containing 5,000 comments from both 2016 and 2022. For subreddits where 2,500 comments from either category did not exist, we sampled the minimum of the number of comments in the two categories from both. We then used the BERT models trained by~\citet{bao_conversations_2021} on Reddit data to assign quantitative supportiveness, agreement, and politeness scores to these comments. All scores were then normalized to fall within the range~$[0,1]$.

We measured the Variance Inflation Factor (VIF) between these metrics in order to determine the degree of correlation between them. For 2016, we obtained VIFs of 4.44, 2.82, and 2.12 for supportiveness, agreement, and politeness respectively and for 2022, we obtained VIFs of 4.96, 3.28, and 2.13 respectively, indicating moderate correlation. To handle this moderate correlation, we followed prior research~\cite{bao_conversations_2021, voigt2017language} to synthesize a single-dimensional prosociality measure. We did this by performing Principal Component Analysis (PCA)~\cite{wold1987principal} over the three metrics, capturing between 75\% and 82\% of the variance in the original data across the different subreddits. 
We then quantified the relationship between this composite prosociality score and a comment's score by regressing the binary upvote label (``high'' or ``low'') on the synthesized prosociality metric, separately for $\dataset{2016}$ and $\dataset{2022}$.

\para{Findings:} Figure \ref{fig:logreg_prosociality} presents the odds ratios from the logistic regression plotted against the number of subscribers for each subreddit that yielded a statistically significant result ($\alpha=0.05$) across each year. The odds ratios represent the likelihood that higher prosociality scores are associated with comments in the ``high'' upvote category, relative to the ``low'' upvote category, for each subreddit.

From the vertical axis, we observe that only four subreddits in total—\texttt{r/aww}, \texttt{r/pokemongo}, \texttt{r/science} in $\dataset{2016}$, and \texttt{r/photoshopbattles} in $\dataset{2022}$—exhibit a \textbf{high odds ratio} of greater than 5. This suggests that in these subreddits, content that is more prosocial is five times more likely to receive a ``high'' number of upvotes.
Additionally, eleven and eight other subreddits in $\dataset{2016}$ and $\dataset{2022}$ respectively show a \textbf{moderate odds ratio} (i.e., greater than one), indicating that higher prosociality is associated with higher likelihood of being highly upvoted, albeit to a lesser degree. 
However, the vast majority of all subreddits (76.3\% in $\dataset{2016}$ and 82\% in $\dataset{2022}$) are clustered at the bottom of the plot with a \textbf{low odds ratio} of less than one. This indicates that across these communities, higher prosociality in content is actually inversely related to the likelihood of being highly upvoted, suggesting that prosocial behavior may not always align with what these communities encourage with upvotes. 


\subsection{Recall of Current Prosociality Measures}

\begin{table}
\small
\sffamily
\centering
  \begin{tabular}{c >{\centering\arraybackslash}p{4cm} c c}
    \multirow{2}{*}{\textbf{Scale}} & \multirow{2}{4cm}{\centering \textbf{Proportion of ``High'' Prosociality Comments}} & \multicolumn{2}{c}{\textbf{\# Values}} \\
    & & 2016 & 2022\\
    \midrule
    \multirow{3}{*}{Macro} & $< 0.1$ & 1 & 0 \\
    & $\geq$ 0.1 $\&$ $<$ 0.25 & 2 & 2 \\
    & $\geq$ 0.25 & 0 & 0 \\
    \midrule
    \multirow{3}{*}{Meso} & $< 0.1$ & 2 & 1 \\
    & $\geq$ 0.1 $\&$ $<$ 0.25 & 8 & 16 \\
    & $\geq$ 0.25 & 3 & 3 \\
    \midrule
    \multirow{3}{*}{Micro} & $< 0.1$ & 20 & 19 \\
    & $\geq$ 0.1 $\&$ $<$ 0.25 & 16 & 23 \\
    & $\geq$ 0.25 & 11 & 7 \\
    \bottomrule
  \end{tabular}
  \caption{Proportion of comments exhibiting macro, meso, and micro values that are also labeled with ``high'' prosociality by existing computational models. Only 14 values from $\dataset{2016}$ and 10 values from $\dataset{2022}$ had more than 25\% recall, while 23 values from $\dataset{2016}$ and 20 values from $\dataset{2022}$ had less than 10\% recall. This demonstrates that existing models lack recall on most of the values extracted by our framework.}
  \label{tab:prosocial_recall}
\end{table}

\para{Methodology:} To conduct an analysis of how well current measures are able to capture the different community values we extracted, we use the 16,000 comments in $\mathcal{D}_{\text{study}}$ from which we computationally extracted values and obtained the PCA-synthesized prosociality scores for these comments using the BERT models as described in the previous section. Next, we threshold these PCA prosociality scores at one standard deviation above the mean, and labeled the comments with a prosociality score above and below this threshold with binary ``high'' and ``low'' prosociality.
Finally, we calculate the proportion of comments that exhibit each of the extracted values and also have a ``high'' prosociality label as determined by the BERT scores, i.e., the \textit{recall of prosociality measures for that particular value}.

\para{Findings:} Table \ref{tab:prosocial_recall} presents the proportion of comments exhibiting extracted values (at macro, meso, and micro scales) and having a ``high'' prosociality label. We observe that there were no macro values assigned to ``high'' prosociality at least 25\% of the time, though there were two macro values each in $\dataset{2016}$ (\textit{relatable} and \textit{concise}) and $\dataset{2022}$ (\textit{relatable} and \textit{humor}) for which between 10\% and 25\% of the pool of comments exhibiting such values were ``high'' prosociality. For the remaining macro value in $\dataset{2016}$ (\textit{humor}), less than 10\% of the comments exhibiting them were labeled with ``high.''

The five values with the highest average recall across $\dataset{2016}$ and $\dataset{2022}$ were \textit{efficiency} (60\%), \textit{prosociality} (46.1\%), \textit{agreement} (43.4\%), \textit{resource management} (33.33\%), \textit{passion} (32.8\%). On the other hand, there were eleven values for which the prosociality measures had 0\% recall in $\dataset{2016}$ and six such values in $\dataset{2022}$.

Overall, we find that there are only 14 and 10 values in $\dataset{2016}$ and $\dataset{2022}$ respectively where the recall of the prosociality measures is at least 25\%. We also found that 26 values from $\dataset{2016}$ and 41 values from $\dataset{2022}$ were captured by the prosociality measures with 10\% and 25\% recall respectively. Finally, there were 23 values from $\dataset{2016}$ and 20 remaining values from $\dataset{2022}$ where the recall is below 10\%, highlighting that existing general-purpose prosociality measures are inadequate at capturing the values being upvoted in practice.

\section{RQ3: Comparison with Prior Taxonomies}\label{sec:prior-taxonomies}

We now compare our computationally-extracted values with prior taxonomies of community and moderator values developed by \citet{weld_making_2024} and \citet{lambert_positive_2024}. We manually inspected all computationally-extracted values and compared each to the categories presented in the two taxonomies from prior work. Through this analysis, we find that the categories in these taxonomies are well represented by our macro, meso, and micro values.

\para{Findings:} Our set of values overlaps with four of the nine categories in the taxonomy of values developed from a survey of Reddit users~\cite{weld_making_2024}. 
These four categories are ``Quality of Content,'' ``Community Engagement,'' ``Participation \& Inclusion,'' and ``Trust,'' which show up as the computationally-extracted values of \textit{personal}, \textit{educational}, \textit{quality}, \textit{engagement}, \textit{inclusivity}, \textit{accuracy}, and \textit{authenticity}. One category not covered in our values was ``norm-adherence,'' which is not detectable from the comment alone without access to community-specific rules. Other categories not present in our set of values include ``Size,'' ``Technical Features,'' ``Diversity,'' and ``Mods.'' These categories are more platform- and demographic-specific which are more high-level and cannot be detected based on the comment text alone. Overall, our set of values covers 11 of the 29 subcategories in this taxonomy, and excluding the five categories not detectable from comment text, our values cover 11 out of 18 subcategories.

The seven uncaptured subcategories from \cite{weld_making_2021} include three related to ``Norms'' (the model lacked subreddit rules), two pertaining to ``Quality of Content'' (model required post metadata), and two under ``Participation and Inclusion'' (model needed user demographics and community-specific tools information). All of these uncaptured subcategories require access to information that is not available from comments alone.

We also find substantial overlap between our extracted values and the taxonomy of attributes that Reddit moderators want to encourage~\cite{lambert_positive_2024}. Specifically, both the taxonomy and our set of values include ``Prosociality,'' ``Participation,'' ``Quality,'' and ``Community-Specific Content.'' Various attributes of prosociality (e.g., \textit{prosociality}, and \textit{respectful}) appear in our set of values as meso and micro values. Similarly, we identify \textit{engagement} as a meso value.
Overall, 
we were able to computationally extract values that map to 14 of the 21 attributes in this taxonomy. The seven attributes not captured are ``Creating a Safe Environment,'' ``Supporting Newcomers,'' ``General Participation,'' ``Norm-Abiding,'' ``Format and Type of Contribution,'', ``External Participation,'' and ``Asking Questions''---several of which cannot be effectively captured by just considering the comment itself. For example, ``Supporting Newcomers'' cannot be inferred without knowing if a comment is replying to a newcomer, and ``General Participation'' and ``External Participation''
are not discernible without augmenting the data provided to the LLM.

\section{Discussion and Implications}

We now discuss the implications of our findings. 

\para{Implications for Online Moderation:} Our findings revealed \textit{macro values} being rewarded consistently across communities through the use of upvotes.
These values represent the traits that most users would like to see from content on Reddit.
\textit{Macro values we identified can be used as ``sensible defaults'' for desirable behavior that moderators could incorporate into their subreddits' guidelines.}

Our analysis also uncovered \textit{meso} and \textit{micro} values rewarded by a smaller set of communities which can often depend on community norms.
For example, the micro value \textit{guidance} which contains extracted keywords such as `constructive advice' and `constructive suggestion' is explicitly discouraged in communities like \texttt{r/depression} and \texttt{r/Showerthoughts} which include rules like
\textit{``No achievement/advice posts or AMAs,''} \textit{``Do not give or request clinical advice or advocate for or against treatments \& self-help strategies,''} and \textit{``No `life pro-tips' or advice''}. In line with these explicit dismissals of advice-giving, we find that \textit{guidance} was not a value extracted from these two communities. Similarly, the meso value \textit{personal} that involves `personal opinion' and `personal experience' are explicitly discouraged in subreddits like \texttt{r/legaladvice} and \texttt{r/science} with rules such as \textit{``Personal anecdotes are off-topic.''} and \textit{``Non-professional personal anecdotes will be removed''} respectively, and we find that \textit{personal} was not a value that we extracted from these communities.
\textit{Therefore, we encourage moderators to explicitly state what is considered to be desirable behavior given that the values we extract seem to be impacted by community guidelines.}

Our work has further implications for Reddit moderators given that prior work has shown that users and moderators do not always agree on the importance of various values~\cite{weld_what_2022}. By understanding preferences and values of their community, moderators can strategically highlight or promote content that aligns with community values for norm-setting and increasing desirable contributions~\cite{kiesler_regulating_2012}. Moreover, our work also reveals that in practice, values that community members reward may evolve over time, which should be taken into consideration for norm-setting.
\textit{Our framework can help moderation teams understand the values their communities are inherently rewarding through positive feedback.}

\para{Improving Models of Desirable Behavior:} Our work highlights a clear distinction between notions of prosociality and content desirability.  
We find that current prosociality measures (e.g., supportiveness, agreement, and politeness) were unable to reliably capture the macro, meso, and micro values we extracted.
Moreover, prosociality metrics alone are inadequate at explaining the relationship between desirable behavior and upvoting patterns on Reddit. 

Different communities hold distinct values and view prosocial behavior in diverse ways, reflecting varying community cultures and norms. Consequently, certain communities may have unique interpretations of what constitutes valuable or upvote-worthy content, which may not always align with traditional prosociality measures. This underscores the need for a deeper exploration into the specific attributes or values that individual communities consider to be desirable.
The misalignment between the values being rewarded in practice and existing computational models of prosociality emphasize the need for better approaches to detect desirable behavior. 
\textit{This calls for the development of models that are tailored towards capturing community-specific values in order to improve recall and align such measures of desirability with the content and behavior that is actually encouraged in practice.}

Being able to detect these values in practice can benefit moderators interested in encouraging desirable behavior in their communities by enabling automated tools to aid in the discovery~\cite{choi_convex_2023} of desirable content. For example, prior work has developed a tool leveraging community-specific models of norms to assist moderators in identifying content to remove~\cite{chandrasekharan_crossmod_2019}. Similarly, modeling desirability can facilitate the process of moderators and users discovering content they want to encourage.

Recent work has explored LLM-based approaches for detecting undesirable content at scale, such as language models for content moderation~\cite{zhan2025slm,goyal2025momoe}. We argue that an important complementary direction is adapting similar frameworks for the identification and \textit{highlighting of desirable content}. Along these lines, \citet{goyal2025language} develop a causal understanding of the linguistic drivers of positive community feedback on Reddit, with \citet{lambert2025mind} applying this to the moderation queue to enable positive reinforcement. Parallel work has also analyzed community-specific models for highlighting desirable content~\cite{goyal2026vastu}. Together with our work, these efforts point toward a more complete picture of computational approaches that can actively surface what communities value.

\para{Extracting Community Values Using LLMs:} Our work provides a simple but powerful framework that uses LLMs to extract community-specific values. Given the scale of Reddit communities, it is not feasible to qualitatively gain insights about each of their perspectives on desirability. Our framework is able to accomplish this task at a much larger scale than qualitative work, while also capturing many of the same insights. A key advantage of our approach is also that values are extracted in a context-dependent manner as notions of what users in online communities find desirable can be highly context dependent. 

Furthermore, the values we extract sufficiently capture content-specific values from prior survey-based taxonomies, primarily at the macro and meso levels. Additionally, we uncover values at the meso and micro levels that are not found in existing taxonomies, thereby extending our understanding of what is considered desirable across Reddit communities.
\textit{Our framework can supplement qualitative work with large-scale analyses of online communities and provide moderators with resources to learn about their communities.}




\section{Limitations and Future Work}
Next, we discuss limitations and directions for future work.

\para{Other Components of Desirability, Lack of Causal Insights:} We focused on upvotes as a proxy for content desirability because of their widespread usage by users and moderators on Reddit.
However, upvotes alone may not capture the full spectrum of desirable behavior in online communities. Other signals such as flairs, awards, and stickied posts are more sparse, but may provide alternative indicators of what communities value. For example, gilded or awarded content may reflect contributions that are more insightful, humorous, or helpful, while stickied posts may signal content of high importance or relevance within a particular community. Moreover, since \texttt{score} is a measure of net-upvotes, our pipeline may have missed the category of \textit{controversial} comments with high upvotes but roughly equal number of downvotes. Future work should incorporate additional signals of desirability to reveal more nuanced patterns of desirable behavior. In addition, our work focuses on comment data from Reddit, and as discussed in Section~\ref{sec:prior-taxonomies}, extraction of other kinds of values requires going beyond comment-level features to subreddit-level metadata and the posts to which these comments replied. Future research can augment our framework with these additional forms of data to extract additional relevant values. Finally, our study contributes a descriptive analysis of values present in highly-scored Reddit comments and does not draw causal insights or provide explanations for \textit{why} certain content got upvoted over others, but rather \textit{describes} the values present within content that was considered to be desirable (i.e., well-received) by online communities.

\para{Cross-platform Study of Desirable Behavior:} Our framework is intended to provide community-specific insights, thus our set of macro, meso, and micro values may not directly apply to communities on other platforms. However, since out framework does not specifically depend on Reddit data in any manner, our methodology can easily be extended to other platforms like YouTube and $\mathbb{X}$ using their specific mechanisms for content engagement (e.g., likes, retweets, YouTube’s ``Super Chat'' and ``Creator Hearts''). Researchers can conduct similar analyses on other platforms to compare the importance of extracted values and reveal how notions of desirability may vary across platforms. Moreover, as AI-populated platforms such as Moltbook emerge, where LLM-driven agents interact in community settings and exhibit meaningful variation compared to human communities~\cite{goyal2026social}, understanding and highlighting desirable contributions in these new sociotechnical environments presents an important open direction.

\para{Use of Closed Source Pre-trained Model:} We relied on GPT-4o model to extract community values encoded in highly-upvoted comments. However, this is a closed-source, pre-trained model, which introduces limitations such as the lack of transparency and potential biases from proprietary training data. This may limit its accessibility for broader use in academic or open research environments. However, due to the lack of studies aimed at computationally extracting values from community-approved desirable content, there were no suitable open-source datasets for fine-tuning on tasks like community value extraction. Our work can be extended by curating appropriate datasets in a similar unsupervised fashion to fine-tuned smaller, open-sourced language models. 

\section{Conclusion}

This paper advances our understanding of desirable behavior in online platforms through a large-scale empirical analysis of highly-upvoted content on Reddit. We present a computational framework that employs large language models (LLMs) to extract community values in a context-dependent manner. Using this framework, we analyze highly-upvoted content from 80 popular subreddits and identify values at three scales: macro (rewarded by most subreddits), meso (rewarded by many but not all subreddits), and micro (rewarded by specific types of subreddits). We highlight that existing prosociality measures alone do not fully explain what types of content get upvoted by communities, and demonstrate the low recall of preexisting computational models to measure prosociality. This underscores the need for more community-specific models capturing the values that are encouraged in practice. Our findings show alignment with prior taxonomies at macro and meso scales, while also providing new insights at the meso and micro scales. This paper lays the groundwork for future exploration into different aspects of desirable behavior online and highlights the need to improve the recall of computational approaches to identify and predict desirable behavior online.

\section*{Ethics Statement}

The data used in this study is comprised entirely of publicly available information. We ensured to rigorously adhere to Reddit's terms of service and privacy policy throughout our data collection and analysis process. Since we utilized a large language model for extracting values from data, we ensured to mask usernames in the comments with the `[NAME]' string and no author information was available to the model. Furthermore, we also ensured to comply with OpenAI's terms of use policies. While our work shows the potential of our framework to extract what values users reward in practice, malicious actors could use a similar approach to identify and promote unwanted behavior and therefore must be used with caution.

\section*{Acknowledgments}
This research was supported by NSF CAREER IIS-2439433. A.G. was supported by compute credits from a Cohere For AI Research Grant and the OpenAI Researcher Access Program. This work used the Delta system at the National Center for Supercomputing Applications through allocation \#240481 from the Advanced Cyberinfrastructure Coordination Ecosystem: Services \& Support (ACCESS) program, which is supported by National Science Foundation grants \#2138259, \#2138286, \#2138307, \#2137603, and \#2138296.

\fontsize{9pt}{8pt} {\selectfont
\bibliography{aaai2026}}


\appendix

\section*{Paper Checklist}

\begin{enumerate}

\item For most authors...
\begin{enumerate}
  \item  Would answering this research question advance science without violating social contracts, such as violating privacy norms, perpetuating unfair profiling, exacerbating the socio-economic divide, or implying disrespect to societies or cultures?
    \answerYes{Yes, online content moderation has for long been an important problem that impacts users online and proactive moderation has been shown to lead to positive outcomes in practice.}
  \item Do your main claims in the abstract and introduction accurately reflect the paper's contributions and scope?
    \answerYes{Yes.}
  \item Do you clarify how the proposed methodological approach is appropriate for the claims made? 
    \answerYes{Yes, justification has been provided throughout the paper.}
  \item Do you clarify what are possible artifacts in the data used, given population-specific distributions?
    \answerYes{Yes, see the ``List of Subreddits in This Study'' section in Appendix \ref{app:subreddit-list}.}
  \item Did you describe the limitations of your work?
    \answerYes{Yes, see the ``Limitations and Future Work'' section in the main text.}
  \item Did you discuss any potential negative societal impacts of your work?
    \answerYes{Yes, see the ``Ethics Statement'' above.}
  \item Did you discuss any potential misuse of your work?
    \answerYes{Yes, see the ``Ethics Statement'' above.}
  \item Did you describe steps taken to prevent or mitigate potential negative outcomes of the research, such as data and model documentation, data anonymization, responsible release, access control, and the reproducibility of findings?
    \answerYes{Yes, see the ``Ethics Statement'' above for data anonymization, ``Code and Data Availability'' section in Appendix \ref{app:code-data-availability} for responsible release, and ``Model and Inference Parameters'' section in Appendix \ref{app:model-parameters} for reproducibility of findings.}
  \item Have you read the ethics review guidelines and ensured that your paper conforms to them?
    \answerYes{Yes.}
\end{enumerate}

\item Additionally, if your study involves hypotheses testing...
\begin{enumerate}
  \item Did you clearly state the assumptions underlying all theoretical results?
    \answerNA{NA}
  \item Have you provided justifications for all theoretical results?
    \answerNA{NA}
  \item Did you discuss competing hypotheses or theories that might challenge or complement your theoretical results?
    \answerNA{NA}
  \item Have you considered alternative mechanisms or explanations that might account for the same outcomes observed in your study?
    \answerNA{NA}
  \item Did you address potential biases or limitations in your theoretical framework?
    \answerNA{NA}
  \item Have you related your theoretical results to the existing literature in social science?
    \answerNA{NA}
  \item Did you discuss the implications of your theoretical results for policy, practice, or further research in the social science domain?
    \answerNA{NA}
\end{enumerate}

\item Additionally, if you are including theoretical proofs...
\begin{enumerate}
  \item Did you state the full set of assumptions of all theoretical results?
    \answerNA{NA}
	\item Did you include complete proofs of all theoretical results?
    \answerNA{NA}
\end{enumerate}

\item Additionally, if you ran machine learning experiments...
\begin{enumerate}
  \item Did you include the code, data, and instructions needed to reproduce the main experimental results (either in the supplemental material or as a URL)?
    \answerYes{Yes, see ``Code and Data Availability'' in Appendix \ref{app:code-data-availability}.}
  \item Did you specify all the training details (e.g., data splits, hyperparameters, how they were chosen)?
    \answerNA{NA}
  \item Did you report error bars (e.g., with respect to the random seed after running experiments multiple times)?
    \answerNA{NA}
  \item Did you include the total amount of compute and the type of resources used (e.g., type of GPUs, internal cluster, or cloud provider)?
    \answerYes{Yes, see ``Model and Inference Parameters'' section in Appendix \ref{app:model-parameters}.}
  \item Do you justify how the proposed evaluation is sufficient and appropriate to the claims made? 
    \answerYes{Yes, see the ``Methodology'' subsections part of the ``RQ2: Examining the Sufficiency of Existing Prosociality Measures in Explaining Upvotes'' section in the main text.}
  \item Do you discuss what is ``the cost`` of misclassification and fault (in)tolerance?
    \answerNA{NA}
  
\end{enumerate}

\item Additionally, if you are using existing assets (e.g., code, data, models) or curating/releasing new assets, \textbf{without compromising anonymity}...
\begin{enumerate}
  \item If your work uses existing assets, did you cite the creators?
    \answerYes{Yes, the original creators of models were cited in the ``Methodology'' subsections of the appropriate section as well as in the ``Related Work'' sections.}
  \item Did you mention the license of the assets?
    \answerNA{NA}
  \item Did you include any new assets in the supplemental material or as a URL?
    \answerNA{NA}
  \item Did you discuss whether and how consent was obtained from people whose data you're using/curating?
    \answerNA{NA}
  \item Did you discuss whether the data you are using/curating contains personally identifiable information or offensive content?
    \answerYes{Yes, see the ``Ethics Statement'' above and ``When are “N/A”s returned by the model?'' subsection in the ``Discussion and Implications'' section in the main text.}
  \item If you are curating or releasing new datasets, did you discuss how you intend to make your datasets FAIR (see \citet{fair})?
    \answerNA{NA}
  \item If you are curating or releasing new datasets, did you create a Datasheet for the Dataset (see \citet{gebru2021datasheets})? 
    \answerNA{NA}
\end{enumerate}

\item Additionally, if you used crowdsourcing or conducted research with human subjects, \textbf{without compromising anonymity}...
\begin{enumerate}
  \item Did you include the full text of instructions given to participants and screenshots?
    \answerNA{NA}
  \item Did you describe any potential participant risks, with mentions of Institutional Review Board (IRB) approvals?
    \answerNA{NA}
  \item Did you include the estimated hourly wage paid to participants and the total amount spent on participant compensation?
    \answerNA{NA}
  \item Did you discuss how data is stored, shared, and deidentified?
    \answerNA{NA}
\end{enumerate}

\end{enumerate}

\section{Data Availability}\label{app:code-data-availability}

To ensure responsible usage of data, we invite researchers interested in using the data to contact the authors for access.

\section{Model and Inference Parameters}\label{app:model-parameters}

Since our framework utilizes a large language model for extracting relevant community values, we state our model specification and inference parameters to facilitate reproducibility. We use OpenAI's GPT-4o~\cite{achiam2023gpt} (\texttt{gpt-4o-2024-08-06}) and 4-bit quantized Gemma-3-27B-IT~\cite{team2025gemma} with $\textit{temperature}=0.0$, and $\textit{top-p}=1.0$. We set the temperature to $0$ to ensure that the extracted values stay consistent across runs for the same \{comment, subreddit, prompt\}, and allow sampling over the entire vocabulary ($\textit{top-p}=1$) to allow the model to output a wide range of values, which we refine later.

The value extraction procedure using GPT-4o including input and output token cost and the generation of explanations for the $16,000$ comments in our dataset cost us approximately $50$ USD, and was performed on an internal cluster using CPUs only. The experiments on Gemma-3-27B-IT was performed on an internal GPU cluster consisting of 3 NVIDIA A40 GPUs.

\section{Prompt Design}\label{sec:prompt-design}

In order to arrive at the prompt we used in our study, we tested a few different prompts and finally settled on the following prompt:

\begin{quote}
    \small\tt ``You are a user on Reddit browsing comments on r/\{SUBREDDIT\}.
    
    Here is the description of the subreddit: \{SUBREDDIT\_DESCRIPTION\}
    
    Below is a comment from a Reddit conversation in r/\{SUBREDDIT\} that received a high number of upvotes. You will also be given the context of the conversation (preceding comment) to the comment in question.
    
    Answer in a brief comma-separated list with either one, two, or three keywords that summarizes certain traits or qualities within the text that might have led to the comment being highly upvoted. If you think only one or two keywords are necessary, then only include that many in your response.
    
    If for some reason you think this comment should have instead been downvoted, answer with 'N/A'.
    
    Perform your analysis in two steps:
    ----------------------------
    
    1. Thinking: Think step by step about the comment, context, and values the comment portrays that might have led to the comment being highly upvoted.
    
    2. Answer: Answer with the brief (MAXIMUM 3) comma-separated list of values that the comment portrays or 'N/A' if the comment should have been downvoted.
    
    ----------------------------
    
    Try to use the same words for related values, trying to reuse the values you have already output for previous comments. For example, if you think 'brevity' is a portrayed value but you already used 'concise', reuse them when appropriate.
    
    Here's the bank of values you have already used:
    \{VALUE\_BANK\}
    
    Note: If a comment contains a line starting with "\&gt;", it means that that line of the comment is a quote from the context, so focus on the part that is not quoted for the values you extract.
    
    Here is the context and comment:
    
    Context: \{CONTEXT\}
    
    Comment: \{COMMENT\}
    ''
\end{quote}

We ask the model to produce its output in strict JSON format for ease of downstream analysis.

\para{Older Prompts} Before settling on the chain-of-thought reasoning prompt above, we also tried various simpler versions like the one below that uses a zero-shot prompting approach, and iteratively modified the prompt upon manual validation to improve the quality of value extraction.

\begin{quote}
    \small\tt ``You are a user on Reddit browsing comments on r/\{$\mathcal{S}_j$\}. Below is a comment from a Reddit conversation in r/\{$\mathcal{S}_j$\} that received a high number of upvotes. \\
    Answer in a brief comma-separated list with either one, two, or three keywords that summarizes certain traits or qualities within the text that might have led to the comment being highly upvoted. If you think only one or two keywords are necessary, then only include that many in your response. \\ 
    If for some reason you think this comment should've instead been downvoted, answer with `N/A'.\\
    Here's the comment:                 
    \{$c^{j}_i$\}\\
    Your list:''
\end{quote}
where $c^{j}_i$ is comment $i$ from subreddit $\mathcal{S}_j$.

\section{Detailed Examples of Extracted Community Values at Different Scales}\label{app:value-list}

We list here the set of micro, meso, and micro values we extract through our methodology. Note that these are not an exhaustive set of values but rather the values that we obtained from our analysis using out framework on the 80 subreddits used in this study.

\subsection{Values for 2016 Comments}

\para{\textbf{Macro Values:}} \textit{humor, relatable, concise}

\para{\textbf{Meso Values:}} \textit{community oriented, prosociality, helpful, nostalgia, clarity, empathetic, engagement, creativity, social commentary, passion, pragmatism, critical perspective, respectful}

\para{\textbf{Micro Values:}} \textit{agreement, insightful, relevance, expertise, resilience, emotional response, objectivity, rationality, growth mindset, authenticity, inclusivity, responsibility, inquisitiveness, encouraging, optimism, ethical, education, gameplay, guidance, safety, well-being, resource management, consistency, efficiency, consideration, historical context, personal, awareness, discourse, controversial, traditionalism, accuracy, shared experience, straightforward, thoughtfulness, strategic thinking, initiative, quality, liberty, cultural appreciation, autonomy, openness, identity, creative expression, entertaining, progressivism, futurism}

\subsection{Values for 2022 Comments}

\para{\textbf{Macro Values:}} \textit{humor, relatable}

\para{\textbf{Meso Values:}} \textit{helpful, respectful, prosociality, critical perspective, concise, nostalgia, accuracy, clarity, engagement, insightful, encouraging, personal, empathetic, inquisitiveness, responsibility, social commentary, authenticity, pragmatism, emotional response, creativity}

\para{\textbf{Micro Values:}} \textit{expertise, community oriented, passion, discussion, liberty, confidentiality, optimism, discourse, straightforward, diversity, historical context, relevance, rationality, inclusivity, efficiency, futurism, well-being, openness, autonomy, philosophical, creative expression, wholesomeness, safety, resilience, dissent, progressivism, systemics, strategic thinking, ethical, guidance, media awareness, sociocultural, traditionalism, initiative, shared experience, resource management, growth mindset, objectivity, politics, entertaining, education, identity, explicit, controversial, thoughtfulness, quality, consumerism, cultural appreciation, agreement}

\section{Further Details on Data Preprocessing}\label{app:thresholds}

\begin{figure}
    \centering
    \includegraphics[width=\linewidth]{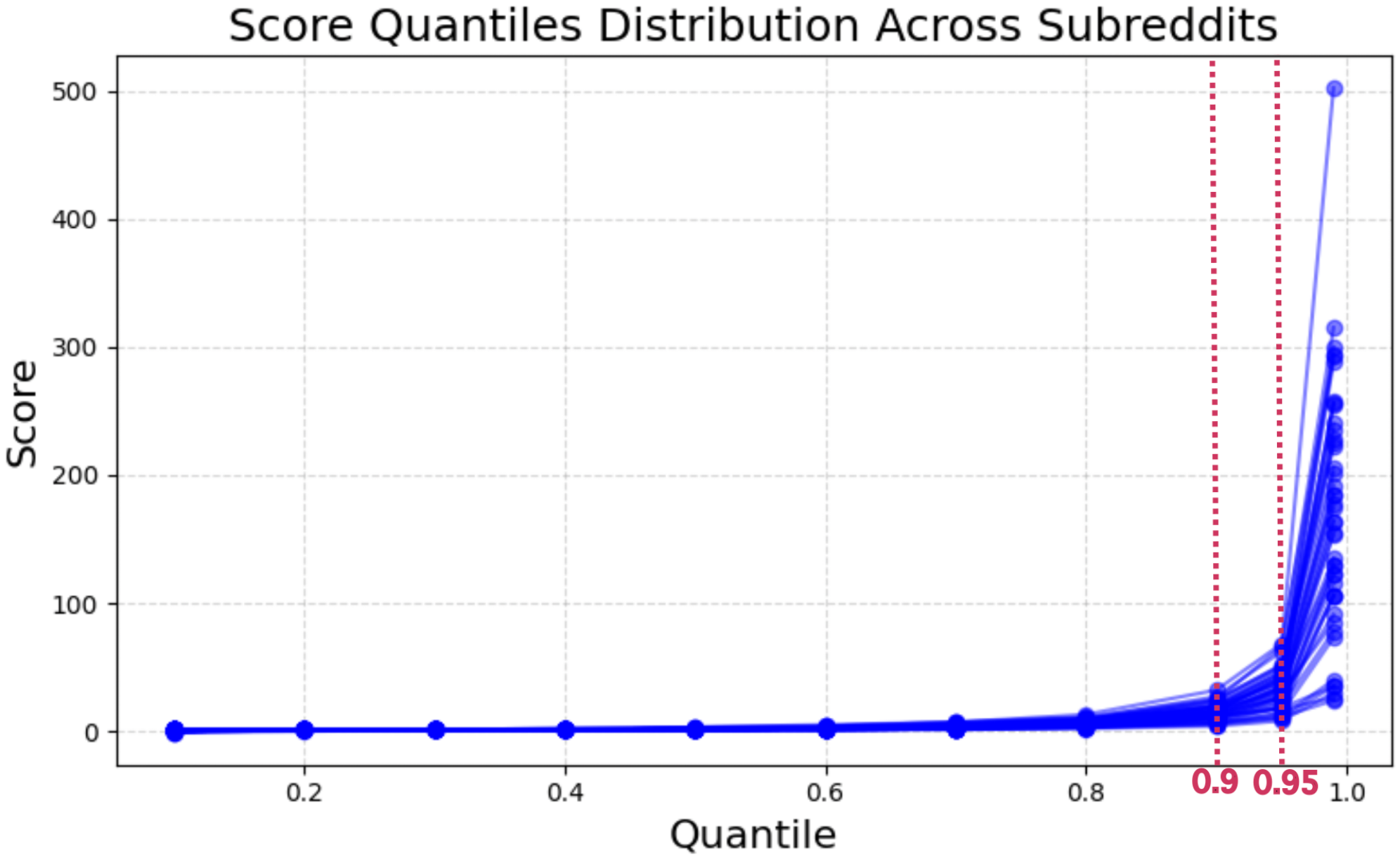}
    \caption{Plot depicting quantiles of the \texttt{score} and two thresholds marked in red at $0.9$ and $0.95$. The first significant rise in the score occurs at $0.95$ which we therefore use as the threshold for ``high'' upvote comments. Since there is little to no rise until $0.7$, we use that as our threshold for ``low'' upvote comments.}
    \label{fig:threshold_upvotes}
\end{figure}

In order to determine thresholds for ``high'' and ``low'' upvote labels, we plotted the quantiles of the \texttt{score} (\# upvotes -- \# downvotes) for each subreddit. From Figure \ref{fig:threshold_upvotes} we notice that a significant rise (elbow) in the score appears at the $95^\text{th}$ percentile across almost all subreddits, and since we are interested in analyzing the highly rewarded content and extract values from them, we use this as our threshold for the ``high'' upvote label. Furthermore, for the logistic regression analysis, we also needed a ``low'' upvote labeled class. For this, we chose $0.7$ as our threshold since most subreddits saw a very minor gradual increase in the score up to the  $70^\text{th}$ percentile indicating that most comments in this bracket received relatively low upvotes.

\section{List of Subreddits in This Study}\label{app:subreddit-list}

We studied $80$ popular subreddits in this work listed below:\\

r/2007scape, r/Android, r/AskHistorians, r/AskTrumpSupporters, r/AskWomen, r/BlackPeopleTwitter, r/CFB, r/CanadaPolitics, r/Christianity, r/DIY, r/DestinyTheGame, r/Futurology, r/Games, r/GetMotivated, r/GlobalOffensive, r/IAmA, r/LateStageCapitalism, r/LifeProTips, r/MMA, r/NSFW\_GIF, r/OldSchoolCool, r/OutOfTheLoop, r/Overwatch, r/PoliticalDiscussion, r/PurplePillDebate, r/SandersForPresident, r/Showerthoughts, r/SubredditDrama, r/TheSilphRoad, r/TwoXChromosomes, r/UpliftingNews, r/anime, r/askscience, r/asoiaf, r/atheism, r/aww, r/books, r/canada, r/changemyview, r/churning, r/creepyPMs, r/dataisbeautiful, r/depression, r/europe, r/explainlikeimfive, r/fantasyfootball, r/food, r/funny, r/gameofthrones, r/gaming, r/gifs, r/hearthstone, r/hiphopheads, r/history, r/india, r/jailbreak, r/legaladvice, r/me\_irl, r/movies, r/nba, r/nosleep, r/nottheonion, r/pcmasterrace, r/personalfinance, r/philosophy, r/photoshopbattles, r/pokemon, r/pokemongo, r/relationships, r/science, r/sex, r/socialism, r/spacex, r/syriancivilwar, r/technology, r/television, r/tifu, r/videos, r/whatisthisthing, r/wow.

\section{Subreddits Where LLM Returns ``N/A''s}\label{app:na-subreddits} The LLM extracted 135 total ``N/A''s coming from a mix of both 2016 and 2022, with 122 comments from 56 communities in 2016 and 13 comments from 12 subreddits in 2022. A list of subreddits returning the ``N/A'' values for each year are listed below.
\para{2016:} r/Android, r/AskHistorians, r/AskTrumpSupporters, r/AskWomen, r/BlackPeopleTwitter, r/CFB, r/CanadaPolitics, r/Christianity, r/DestinyTheGame, r/Futurology, r/Games, r/GetMotivated, r/GlobalOffensive, r/LifeProTips, r/NSFW\_GIF, r/Overwatch, r/PoliticalDiscussion, r/PurplePillDebate, r/SandersForPresident, r/Showerthoughts, r/SubredditDrama, r/UpliftingNews, r/anime, r/askscience, r/atheism, r/aww, r/books, r/canada, r/changemyview, r/churning, r/dataisbeautiful, r/depression, r/europe, r/explainlikeimfive, r/food, r/funny, r/gameofthrones, r/gifs, r/hearthstone, r/history, r/india, r/legaladvice, r/movies, r/nba, r/nosleep, r/nottheonion, r/personalfinance, r/philosophy, r/photoshopbattles, r/pokemon, r/relationships, r/science, r/sex, r/spacex, r/syriancivilwar, r/technology
\para{2022:} r/AskHistorians, r/DIY, r/IAmA, r/LifeProTips, r/UpliftingNews, r/depression, r/hiphopheads, r/nosleep, r/pokemongo, r/sex, r/tifu, r/videos

\section{Replication of Value Extraction with 4-bit Quantized Gemma-3-27B-IT}\label{app:gemma-replication}
To mitigate potential effects of bias due to using a closed source model in GPT-4o, we perform a replication of value extraction using 4-bit quantized version of Gemma-3-27B-IT~\cite{team2025gemma}. Below are the list of values extracted by the LLM for both 2016 and 2022.
\subsection{Values for 2016 Comments}
\para{\textbf{Macro Values:}} \textit{humor, relatable, concise}

\para{\textbf{Meso Values:}} \textit{community oriented, prosociality, helpful, nostalgia, clarity, empathetic, engagement, creativity, social commentary, passion, critical perspective, respectful}

\para{\textbf{Micro Values:}} \textit{agreement, pragmatism, insightful, relevance, expertise, resilience, emotional response, objectivity, rationality, growth mindset, authenticity, inclusivity, responsibility, inquisitiveness, encouraging, optimism, ethical, education, gameplay, guidance, well-being, consistency, efficiency, consideration, historical context, personal, awareness, discourse, accuracy, straightforward, thoughtfulness, strategic thinking, initiative, quality, cultural appreciation, autonomy, openness, entertaining, futurism}
\subsection{Values for 2022 Comments}
\para{\textbf{Macro Values:}} \textit{humor, relatable, concise}

\para{\textbf{Meso Values:}} \textit{helpful, respectful, prosociality, critical perspective, nostalgia, accuracy, clarity, engagement, insightful, encouraging, personal, empathetic, inquisitiveness, responsibility, social commentary, authenticity, creativity}

\para{\textbf{Micro Values:}} \textit{expertise, pragmatism, community oriented, passion, liberty, optimism, discourse, straightforward, diversity, historical context, relevance, rationality, inclusivity, efficiency, futurism, well-being, openness, autonomy, philosophical, wholesomeness, resilience, dissent, progressivism, strategic thinking, ethical, guidance, sociocultural, traditionalism, initiative, shared experience, growth mindset, objectivity, politics, entertaining, education, explicit, controversial, thoughtfulness, quality, cultural appreciation, agreement}

\end{document}